\documentstyle[aps,prl,multicol,fancyheadings]{revtex}

\def\bbbc{{\mathchoice {\setbox0=\hbox{$\displaystyle\rm C$}\hbox{\hbox
to0pt{\kern0.4\wd0\vrule height0.9\ht0\hss}\box0}}
{\setbox0=\hbox{$\textstyle\rm C$}\hbox{\hbox
to0pt{\kern0.4\wd0\vrule height0.9\ht0\hss}\box0}}
{\setbox0=\hbox{$\scriptstyle\rm C$}\hbox{\hbox
to0pt{\kern0.4\wd0\vrule height0.9\ht0\hss}\box0}}
{\setbox0=\hbox{$\scriptscriptstyle\rm C$}\hbox{\hbox
to0pt{\kern0.4\wd0\vrule height0.9\ht0\hss}\box0}}}}
\input psfig.tex
\def\sinn{\mathop{\rm sinn}\nolimits}
\begin{document}
\draft
\preprint{\large \bf DRAFT \now}
\title {Supernovae evidence for an accelerating expansion of the universe}
\author{George A. Baker, Jr.}
\address{
Theoretical Division, Los Alamos National Laboratory\\
University of California, Los Alamos, N. M. 87544 USA }
\date{submitted june 16, 2000}
\maketitle
\begin {abstract}
Recent experimental results find strong indications that the universe is
flat, while other experimental results from supernovae Ia observations
have been interpreted to show that, not only that there is an accelerating
expansion of the universe, but also that the universe is strongly curved.
By means of a recently proposed metric, I am able to show that the
experimental results which had previously been analyzed to give a strong
curvature are quite consistent with a flat universe, thus resolving the
apparent mismatch. The conclusion  of these latter authors which indicated
an accelerating  expansion of the universe is unchanged.

\end {abstract}

\vspace*{-0.4cm}
\begin{multicols}{2}

\columnseprule 0pt

\narrowtext
Recently de Bernardis {\it et al.}\cite{deB} reported that very careful
examination of the cosmic microwave background provides, as further explained
by Hu\cite{Hu}, very strong evidence that the universe is flat!

On the other hand,
data has recently been reported by Riess {\it et al.}\cite{Riess}, and
Perlmutter {\it et al.}\cite{Perl} on measurements of the luminosity and the
redshift of a number of high-redshift supernovae of type Ia.
The authors' best fit to their data involves a strongly curved
spacetime. The curvature constant $\Omega _k$ (defined below) that they find
is about -1.05, which corresponds
to a strongly curved, open universe. These authors have
analyzed their data under the assumption of the Friedman-Lema\^\i tre line
element at large distances and the resulting cosmological relation between
the rate of expansion, the mean mass density, the radius of curvature of
space, and the cosmological constant. This relation follows from Einstein's
field equations.\cite{Peb} The authors have concluded, in terms of this model
of the universe, that instead of the rate of expansion decreasing, as many
workers had thought, their data is best fit by a model in which the rate
of expansion is increasing. Riess {\it et al.}\cite{Riess} find that the
deceleration/acceleration parameter $q_0< 0$ (acceleration) with better
than 90\% confidence, and Perlmutter {\it et al.}\cite{Perl} find the same at
the 2.86 standard deviation level of confidence. Similar error limits
apply to $\Omega _k$.

    I have recently found\cite{Bak} that the commonly used ``Swiss cheese
model,'' in which a static, exterior, Schwarzschild solution is inserted in
an expanding Freidman-Lema\^\i tre metric, possesses the unphysical property
that some of the trajectories are discontinuous functions of their initial
conditions.  This feature is, I think, unacceptable.  I have proposed an
alternate metric which avoids this problem. In this letter, I apply this
alternate metric to the analysis of the above cited supernovae data.  I
find that, the principal conclusion, that an accelerating universe best
fits the data remains unchanged.  However the mean density and cosmological
constant parameters describing the best fit change significantly, and are
now quite consistent with a flat universe. The alternate line element
is,\cite{Bak}
\begin{equation}
ds^2=-e^\mu (d\hat r^2+\hat r^2d\theta ^2+\hat r^2\sin ^2\theta d\phi ^2)
+e^\nu dt^2, \label{1}
\end{equation}
where
\begin{eqnarray}
e^\mu &=&{{a(t)^2}\over {\displaystyle \left [1+\left ({{a(t)\hat r}\over
{2a(t)R}}\right )^2\right ]^2}}\left (1+{m\over {2a(t)\hat r}}
\right )^4,\label{2} \\
e^\nu &=&c^2\left ({{\displaystyle 1-{m\over {2a(t)\hat r}}}\over
{\displaystyle 1+{m\over {2a(t)\hat r}}}}\right )^2 ,\quad
m\equiv {{GM}\over {c^2}}\left [1+\left ({\hat r\over {2R}}\right )^2\right ]
^{1/2} \nonumber
\end{eqnarray}
where $G$ is Newton's constant of gravitation, and $M$ is the mass concentrated
at the origin.  This metric has dynamics which are undetectably different
from those of the Schwarzschild metric in the solar system, and the metric
is asymptotically equal to the Freidmann-Lema\^\i tre metric at great distances.
This metric was developed in the context of gravitationally bound systems,
and, as is the case for the Schwarzschild metric, it was only considered
valid for $m/2r \ll 1$. This being said, it makes physical sense, I think,
for the effects of the long range gravitational force not to be cut-off
at some finite distance well short of the event horizon, as is the case when
the Friedmann-Lema\^\i tre metric is used to analysis high redshift objects.
This view provides a physical justification to use my alternate metric
in this analysis.

In order to apply this result to the present case, we make
the further assumption that, on the scale of the distance to the high redshift
supernovae under discussion, the universe is homogeneous and isotropic.
Birkoff's theorem\cite{Bir} says in a homogeneous, zero pressure cosmology
that outside a sphere, within which the mass is spherically symmetrically
distributed, one finds  that on
the surface of the sphere (and outside as well), the effects are the same as
though the total mass was concentrated at the origin.  This theorem
suggests that when
considering a particular supernova, we can take $M$ to be the total mass
inside the sphere centered on that supernova and which has the observer (us)
on its surface.

We may take as a helpful guide to the proper {\it modus operandi}, Carroll,
Press, and Turner.\cite{CPT} The relation between
the rate of expansion of the universe, given by the universal multiplier
$a(t)$ defined by $\vec r(t)=a(t)\vec r(t_0)/a(t_0)$ is\cite {Bak}
\begin{equation}
{{8\pi G\rho }\over {c^2}}=
8\pi T_4^4= {3\over {\displaystyle a(t)^2R_0^2\left (1+ {m\over {2a(t)\hat r}}
\right )^5}} +{{3\dot a^2}\over {c^2a^2}}-\Lambda , \label{3}
\end{equation}
where $c$ is the velocity of light, $\rho $ is the mean density, $T$ is the
stress-energy tensor, and an overdot means differentiation with respect to $t$.
We define the standard parameters,
\begin{equation}
\Omega _M ={{8\pi G\rho _0}\over{3H_0^2}},\quad \Omega _\Lambda =
{{\Lambda c^2}\over {3 H_0^2}}, \quad \Omega _k=-{{c^2}\over {R_0^2H_0^2}},
\label{4}
\end{equation}
where $H_0=\dot a(t_0)/a(t_0)$ is the Hubble constant at the present time,
$\Lambda $ is the cosmological constant, and $R_0$ is the radius
of curvature at the present time.  Note that as usual, $R_0^2$ is negative
in an open universe.  In these terms, Eq.\ \ref{3} becomes
\begin{equation}
\left ({{\dot a}\over {aH_0}}\right )^2=\Omega _M
\left ({\rho \over {\rho_0}}\right )+{{R_0^2\Omega _k}\over {\displaystyle
R^2\left (1+{m\over {2a\hat r}} \right )^5}} +\Omega _ \Lambda \label{5}
\end{equation}
Since at the present time, the look-back time is, of course, zero, as explained
above the radius of the sphere enclosing total mass $M$ will be zero. Thus,
as the mass inside the sphere vanishes like the cube of the radius,
the term $m/2a\hat r\propto \hat r^2\to 0$. Hence, from Eq.\ \ref{5}, at the
present time, we have
\begin{equation}
1=\Omega _M+\Omega _k+\Omega _\Lambda .\label{6}
\end{equation}
Using this result, we may eliminate $\Omega _k$ and rewrite Eq.\ \ref{5}
as,
\begin{eqnarray}
\left [{{\dot a(t)}\over {H_0}}\right ]^2 &=& \left (1+{m\over {2a(t)\hat r}}
\right )^{-5}  \nonumber \\
&&+\Omega _M\left [{1\over a}-\left (1+{m\over {2a(t)\hat r}}\right )
^{-5}\right ]
\label{7} \\
&&+\Omega _\Lambda \left [a^2-\left (1+{m\over {2a(t)\hat r}}\right )^{-5}
\right ] \nonumber
\end{eqnarray}

The next quantity that we need is the distance as a function of the observed
luminosity and  the redshift for our line element.  As is well known\cite{Peb}
$a(t)/a(t_0)=1/(1+z)$.  Thus from Eqns.\ \ref{1}, \ref{2}, and \ref{7}, we may
deduce
\begin{eqnarray}
\dot z&=&H_0(1+z)\Bigg \{ (1+z)^2\left (1+{m\over{2a\hat r}}\right )^{-5}
\nonumber \\ &&+\Omega _M
\left [ (1+z)^3-(1+z)^2\left (1+{m\over{2a\hat r}}\right )^{-5}\right ]
\label{8} \\
&&+\Omega _\Lambda\left [1-(1+z)^2\left (1+{m\over{2a\hat r}}\right )^{-5}
\right ] \Bigg  \} ^{0.5}. \nonumber
\end{eqnarray}
The proper distance is just $c$ times the proper lookback time which gives
\begin{eqnarray}
e^{0.5\mu }d\hat r&=& e^{0.5\nu}{{dz}\over {\dot z}} \nonumber \\
{{dr}\over {\sqrt{\displaystyle 1-\left ({r\over R}\right )^2}}}&=& c\left [
{{\displaystyle 1-{m\over {2a\hat r}}}\over {\displaystyle \left ( 1+
{m\over {2a\hat r}}\right )^3}}\right ]
{{(1+z)\, dz}\over {\dot z}}, \label {9}
\end{eqnarray}
where we have used the standard change of variables,
\begin{equation}
r={{\hat r}\over{\displaystyle 1+\left ({{\hat r}\over {2R}}\right )^2}} .
\label{10} \end{equation}
If we now integrate Eq.\ \ref{9}, we get, following Carroll {\it et al.},
the lookback time and the luminosity distance,
\begin{eqnarray}
t_0-t_1&=&\int _0^{z_1}\left({{\displaystyle 1-{m\over {2a\hat r}}}\over
{\displaystyle 1+{m\over {2a\hat r}}}}\right ){{dz}\over {\dot z}} \nonumber \\
D_L&=&cH_0^{-1}(1+z)|\Omega _k|^{-0.5} \label{11}\\
&&\times\sinn\left \{|\Omega _k|^{0.5}
\int _0^z \left [{{\displaystyle 1-{m\over {2a\hat r}}}\over {\displaystyle
\left ( 1+{m\over {2a\hat r}}
\right )^3}}\right ] {{(1+z)\, dz}\over {\dot z}}\right \} \nonumber
\end{eqnarray}
where $\dot z$ is given by Eq.\ \ref{8} and ``$\sinn $'' is defined as
$\sinh $ when $\Omega _k> 0$ (open universe) and as $\sin $ when
$\Omega _k <0$ (closed universe). These equations coincide with the
results of Carroll {\it et al.}\cite{CPT}, when $m=0$, as expected.

To complete the computation, it remains to supply $m/2a\hat r$ as a function
of the redshift. For notational convenience we introduce the following
dimensionless variables,
\begin{eqnarray}
{\mathcal R}&=&H_0\hat r/c,\quad {\mathcal M}={{H_0^3M(\hat r)}\over
{c^3\rho _0}}, \nonumber \\
v(z)&\equiv &{m\over{2a\hat r}}\equiv {{GM(\hat r)} \over {2c^2a\hat r}}
\nonumber \\&=&{{3(1+z)\Omega _M{\mathcal M}}\over {16\pi {\mathcal R}}}
\left (1-0.25\Omega _k{\mathcal R}^2\right )^{0.5} \label {12} \\
q(z)&=&\left \{ (1+z)^2+\Omega _M\left [(1+z)^3(1+v)^5-(1+z)^2\right ]\right .
\nonumber \\ &&\left . +\Omega _\Lambda \left [(1+v)^5-(1+z)^2\right ]\right \}
^{0.5} \nonumber
\end{eqnarray}
If $l$ is the proper distance, then by Hubble's law, $z=H_0l/c$.  By use
of Eqs.\ \ref{1}, \ref{2}, \ref{4}, \ref{7}, and Hubble's law, we may deduce the
following two non-linear differential equations,
\begin{eqnarray}
{{d{\mathcal R}}\over {dz}}&=&{{-(1+z)\left (1-0.25\Omega _k{\mathcal R}^2
\right )(1+v)^{0.5}} \over {q(z)}}
\nonumber \\
{{d{\mathcal M}}\over {dz}}&=&{{-4\pi (1+z)(1+v)^{6.5}{\mathcal R}^2}\over {
\left (1-0.25 \Omega _k{\mathcal R}^2\right )^2q(z)}}. \label{13}
\end{eqnarray}
The initial conditions are ${\mathcal R}={\mathcal M}=0,\; z=z_1$ and
we integrate the equations from $z=z_1$ to $z=0$.

\begin{figure}
\vskip 0.5\baselineskip
\centerline {\psfig{figure=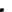,width=\hsize}}
\vskip 0.5\baselineskip
\figure{FIG.\ 1\ \ The metric correction term $v(z)=m(\hat r)/2a\hat r$
for the five models A -- E. The intersection of these curves with the
dotted line corresponds to the special point $z_c$ in each of
these models. The end of each curve corresponds to the last integration
point before $v$ blows up. \label{fig.1}}
\end{figure}

I have solved these equations by the Runge-Kutta method.  To do so I have
rewritten the appropriate routine in Press {\it et al.}\cite{Pr} in double
precision.
Some sample cases, Models A -E\cite{CPT}, are shown in Fig.\ 1.
In this figure I plot the potential, $v(0,z_1)$, through which light must
climb to reach an observer who sees a redshift of $z_1$.
The five models are as follows
Model A, $\Omega _M=1,\; \Omega _\Lambda =0$; Model B, $\Omega _M=0.1,\;
\Omega _\Lambda =0$; Model C, $\Omega _M=0.1,\; \Omega _\Lambda =0.9$;
Model D, $\Omega _M=0.01,\; \Omega _\Lambda =0$; Model E, $\Omega _M=0.01,\;
\Omega _\Lambda =0.99$. These are all either flat or open models.
One will note that the principal effect at small $z_1$ is the mass density.
It is to be further noted that the point $v(z_c)=1$ is special. At this point,
$dD_L/dz$ and $dt/dz$ would change sign, as can be seen in Eq.\ \ref{11}.
In Fig.\ 2 we display the lookback time as a function of $z$
for these same models. The point $z_c$ is marked by a large dot.
It is to be noticed that the lookback time begins to decrease
with $z$ before this special point is reached as a foreshadowing of the
special point.  In addition, ${\mathcal R}(0,z_1)$ also begins to decrease
before the special point is reached, but ${\mathcal M}(0,z_1)$ and $v(0,z_1)$
both increase right through this point and blow up shortly there after,
as indicated by the end of the lines in Figs.\ 1 and 2.  These results are
numerical as the proper analysis of Eq.~\ref{13} is beyond the scope
of this letter.

\begin{figure}
\vskip 0.5\baselineskip
\centerline {\psfig{figure=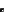,width=\hsize}}
\vskip 0.5\baselineskip
\figure{FIG.\ 2\ \ The lookback time for the five models A -- E.
The large dot on each of these curves correspond to the intersection points
in Fig.\ 1. \label{fig.2}}
\end{figure}

It is not my purpose to obtain a full statistical analysis of the redshift
-- luminosity data of \cite{Riess} and \cite {Perl}. Rather I have worked
with the best fit of \cite{Perl}, $\Omega _M=0.73,\; \Omega _\Lambda =1.32$.
Specifically I find that I can fit
this best fit for the luminosity distance as a function of redshift to
within $\pm 0.02$ magnitudes over the range of the data used.  This is,
I think, at least as close as that curve accurately represents the data.

My best fit is $\Omega _M=0.22\pm 0.03,\; \Omega _\Lambda =0.91\pm 0.15$
for the accuracy quoted above. My best fit for flat space is $\Omega _M=0.19,
\; \Omega _\Lambda =0.81$.  The magnitude of $v(0,z_1)$ is roughly proportional
to $z^2$, and its largest value in both cases is less than $0.07\ll 1$.
When the statistical and systematic errors are taken into account the
uncertainties are probably about 3 to 4 times larger than those quoted above.
My result for flat space is almost the same as Model C of Carroll
{\it et al.}\cite{CPT}. That is to say that it is flat, has a plausible
amount of matter and involves the use of the cosmological constant. Note
is also taken that for the special case of flat space, Eq.\ \ref{7} reduces
to Eq.\ 9 of reference \cite{CPT}; however, our Eq.\ \ref{9} still reflects
a change in the metric.  Our results predict that the acceleration parameter
$q_0\approx -0.79$,
which is strongly negative and therefore agrees with the main conclusion of
references \cite {Riess} and \cite{Perl} that the expansion of the universe
is accelerating.  We have shown however that the disagreement between their
best-fit, curvature results and those of references \cite{deB} and \cite {Hu}
can be resolved by employing my alternate metric.

\end{multicols}

\end{document}